\documentclass[aps,prl,twocolumn,noshowpacs,superscriptaddress,groupedaddress,nofootinbib,preprintnumbers]{revtex4}
\pdfoutput=1

\usepackage{amsmath}
\usepackage{amsfonts}
\usepackage{amssymb}
\usepackage{graphicx, rotating}
\usepackage{epstopdf}
\usepackage{epsfig}
\usepackage{latexsym}
\usepackage{graphicx}
\usepackage{color}
\usepackage{amsmath,amssymb}
\usepackage{slashed}
\usepackage{hyperref}
\hypersetup{colorlinks, citecolor=bluscuro, linkcolor=black, urlcolor=bluscuro}
\definecolor{rossos}{cmyk}{0,1,1,0.55}
\definecolor{bluscuro}{rgb}{0.15, 0.2, .85}
\definecolor{bluchiaro}{cmyk}{1,.3,0.,0.1}

\newcommand{\eq}[1]{Eq.~(\ref{#1})}

\newcommand{\nn}{\nonumber}

\newcommand{\be}{\begin{equation}}
\newcommand{\ee}{\end{equation}}
\newcommand{\bea}{\begin{eqnarray}}
\newcommand{\eea}{\end{eqnarray}}
\newcommand{\bc}{\begin{center}}
\newcommand{\ec}{\end{center}}

\def\lra#1{\overset{\text{\scriptsize$\leftrightarrow$}}{#1}}
\def\hf{\hat h}
\def\h{h}

\def\stw{s_{\theta_W}}
\def\ctw{c_{\theta_W}}
\def\ttw{t_{\theta_W}}

\begin{document}



\title{BSM Primary Effects}

\author{Rick S. Gupta}
\affiliation{IFAE, Universitat Aut{\`o}noma de Barcelona, 08193~Bellaterra,~Barcelona}

\author{Alex Pomarol}
\affiliation{Dept. de F\'isica, Universitat Aut{\`o}noma de Barcelona, 08193~Bellaterra,~Barcelona}

\author{Francesco Riva}
\affiliation{Institut de Th\'eorie des Ph\'enom\`enes Physiques, EPFL,1015 Lausanne, Switzerland}

\begin{abstract}
Using the predictive power of the effective field theory approach, we present a physical parametrization of the leading effects beyond the SM (BSM), that give us at present the best way to constrain heavy new-physics at low-energies. We show that other BSM effects are not independent from these ones, and we provide the explicit correlations. This information is useful to know where to primarily look for new physics in future experiments, and to  know how this new physics is related to previous measurements, most importantly in electroweak-symmetry breaking processes or Higgs physics.
\end{abstract}

\maketitle

The absence at the LHC of new physics beyond the SM  (BSM)
 suggests that the characteristic scale of  new-physics $\Lambda$ must be heavier than the electroweak (EW) scale.
Assuming this, one can obtain an
SM effective-theory  as an expansion in   SM fields and  derivatives over $\Lambda$: 
${\cal L}_{\rm eff}= {\cal L}_4+{\cal L}_6+\cdots$, 
where ${\cal L}_4$ is made of dimension-4 operators and   defines  what we call the SM Lagrangian, while 
${\cal L}_6$, that contains dimension-6 operators,  gives the   leading  BSM effects.\footnote{Assuming lepton and baryon number conservation.} 
While the  predictions  from ${\cal L}_4$   have been fully addressed and tested experimentally,
 it becomes now crucial to understand what are the complete predictions   from ${\cal L}_6$.

This is the main goal of this article.
Our approach  however will  differ from the usual one, in which
one  starts with the  set of independent operators  in  ${\cal L}_6$ \cite{Buchmuller:1985jz,Grzadkowski:2010es}, 
and relates their Wilson coefficients to observables \cite{Han:2004az, Pomarol:2013zra}.
We will follow here a  bottom-up approach, 
our starting point  being all  possible  new interactions  among SM fields that can be 
induced by BSM physics.
 Not all these interactions can be obtained  from ${\cal L}_6$ and, 
of the possible ones, not all of them are independent.
Our  purpose  is to   find  the set of independent  new-interactions that can arise from ${\cal L}_6$
and are, at present, the experimentally best tested ones.
This set of physical quantities, that give us  the best  way to constrain new physics,
will be called  BSM primary effects.
  For electroweak processes  these are, in some sense, a generalization of the well-known $S$ and $T$ parameters~\cite{Peskin:1990zt}.

Our   main result   will be to show which  new-physics effects  are not independent and are instead correlated with the BSM primaries (see also Ref.~\cite{Contino:2013kra}).
 This information can be  useful to understand where to prioritize  new physics searches. Furthermore, in the future, if  a deviation from the SM is measured, 
  the correlations that we present  will  tell us where other BSM effects have to be found.

We start with   the CP-conserving BSM-effects. It is useful to parametrize
these effects as $\hf$-dependent interactions, where
$\hf$ is  the neutral component of the Higgs-field:
\be
 \hf \equiv v+h(x)\, ,
\ee
where $v\simeq 246$ GeV  is the Higgs vacuum expectation value (VEV) and $\h$  the Higgs excitation.
We first consider those effects which  affect  interactions involving only $h$, that 
 arise from dimension-6 operators generated by multiplying  operators in   ${\cal L}_4$
by    $|H|^2/\Lambda^2$ ($H$ being  the Higgs doublet).
In the unitary gauge, that will be used henceforth, 
these effects can be captured by promoting  the SM parameters,
 that we take to be $e,\stw,g_s,Y_{f},\lambda_h$, and the Higgs kinetic-term $Z_h$,
\footnote{The effect of $|H|^2/\Lambda^2$ multiplying the  SM fermion  kinetic-terms is redundant as   can  be eliminated by a redefinition of the SM fermions.}
 to $\hf$-dependent functions: \begin{gather}
e(\hf),\ \stw(\hf),\ g_s(\hf),\ Y_f(\hf),\ \lambda_h(\hf),\ Z_h(\hf)\, .
\label{generals}
\end{gather}
These functions can be expanded in powers of  $\hf^2/\Lambda^2$, {\it e.g.}  $e(\hf)= e+\delta e\, {\hf^2}/{v^2}+\cdots$, where (here and in what follows) we absorb powers of  $v^2/\Lambda^2$ in the expansion coefficients. 
In the vacuum $\hf=v$, \eq{generals} only implies a redefinition of the SM parameters with no impact on physical processes, {\it i.e.},  these effects    can only be probed in Higgs physics. 
To understand the   effects from $e(\hf),\stw(\hf),g_s(\hf)$,
it is convenient 
to write the SM gauge-interactions in a non-canonical way, since this is suitable to accommodate
 space-time dependent couplings while  keeping gauge-invariance manifest. For the EW sector
 this is given by
\begin{align}
\mathcal{L}_{\rm EW}&=-\frac{1}{4e^2(\hf)}\Big( A_{\mu\nu}+\stw^2(\hf)\, Z_{\mu\nu}\Big)^2
-\frac{\ctw^2(\hf)}{4g^2(\hf)}  Z_{\mu\nu}^2\nn\\
&-\frac{1}{2g^2(\hf)} W^+_{\mu\nu}W^{-\mu\nu}
+ \frac{\hf^2}{4}\left[ W_\mu^+ W^{-\mu}+\frac{1}{2}Z^\mu Z_\mu\right]\nn\\
&+
A_\mu J^\mu_{em}+ Z_\mu J^\mu_{3}+  W_\mu^+ J^\mu_{+} + W_\mu^-  J^\mu_{-}\, ,
\label{SMnoncan}
\end{align}
where 
$g(\hf)\equiv e(\hf)/\stw(\hf)$,  and
 we have defined  $J^\mu_{\pm}\equiv(J^\mu_{1}\pm i J^\mu_{2})/\sqrt{2}$, $J_{em}^\mu\equiv J_3^\mu+ J_Y^\mu$, $Z_{\mu \nu}\equiv \hat{Z}_{\mu \nu}-i   W^+_{[\mu}W^-_{\nu]} $, $A_{\mu \nu}\equiv \hat{A}_{\mu \nu}$,  $ W^{\pm}_{\mu\nu}\equiv\hat{W}^{\pm}_{\mu \nu}\pm i  W^{\pm}_{[\mu}(A+Z)_{\nu]} $ with   $\hat{V}_{\mu\nu}\equiv\partial_\mu V_\nu-\partial_\nu V_\mu$,
and $J_a^\mu$ ($a$=$1,2,3$) and $J_Y^\mu$ being respectively the $SU(2)_L$ and $U(1)_Y$  currents. 
In the vacuum $\hf=v$, \eq{SMnoncan} gives the SM EW-interactions  after
 substituting   $ A_\mu\rightarrow A_\mu -\stw^2(v) Z_\mu$ and canonically normalizing the gauge-boson fields.
The two independent deviations w.r.t. the SM,  parametrized by $e(\hf)$ and $\stw(\hf)$, 
can be projected orthogonally into two different physical processes, that  we choose to be $h\to \gamma\gamma$ and $h\to Z\gamma $ for the accuracy to which they are experimentally  constrained. Indeed,
a $h\gamma\gamma$ coupling can arise from the $\hf$-dependence of $e$  with
constant $\stw$:
\be
e(\hf)=e(1+\boldsymbol{\kappa_{\gamma\gamma}}\frac{\hf^2}{{v^2}})\,  ,\ \ \ \ \ \stw(\hf)=\stw\, ,
\label{kggkzg}
\ee
that  plugged into  \eq{SMnoncan}  gives, in the canonical basis, the BSM-terms:
\bea
\Delta {\cal L}^h_{{\gamma\gamma}}&=& \boldsymbol{\kappa_{\gamma\gamma}}  
\left(\frac{ \h}{v}+\frac{\h^2}{2v^2}\right)
\Bigg[ A_{\mu\nu}A^{\mu\nu}\nn\\&&+  Z_{\mu\nu}Z^{\mu\nu}+ 2 W^+_{\mu\nu}W^{-\mu\nu}\Bigg]\, ,
\label{gamgam}
\eea
where now and henceforth  $Z_{\mu \nu}\equiv \hat{Z}_{\mu \nu}-i  g \ctw W^+_{[\mu}W^-_{\nu]} $, $A_{\mu \nu}\equiv \hat{A}_{\mu \nu}-i g  \stw W^+_{[\mu}W^-_{\nu]}$ and $W^{\pm}_{\mu\nu}\equiv\hat{W}^{\pm}_{\mu \nu}\pm i  g W^{\pm}_{[\mu}(\stw A+\ctw Z)_{\nu]} $. The first term of \eq{gamgam} contains a $h\gamma\gamma$ coupling
and  corresponds to    our  first BSM primary effect:  it defines the  best observable
that can be used to bound all terms in $\Delta {\cal L}^h_{{\gamma\gamma}}$.
 Indeed, from the experimental value of $h\to \gamma\gamma$ \cite{Atalscouplings} we obtain
 bounds on  $\boldsymbol{\kappa_{\gamma\gamma}}$ at the per-mille level~\cite{Pomarol:2013zra}. 

On the other hand,  we can take, {\it orthogonally}
to \eq{kggkzg}, the direction
\be\label{delatastw}
e(\hf)=e\, , \ \ \ \ \   
{s^2_{\theta_W}(\hf)}=s^2_{\theta_W}(1-\boldsymbol{\kappa_{Z\gamma}}\frac{\hf^2}{{ v^2}})\, ,
\ee
that in \eq{SMnoncan} gives the  BSM-induced interactions
\bea
\Delta {\cal L}^h_{{Z\gamma}}&=&\boldsymbol{\kappa_{Z\gamma}} \left(\frac{\h}{v}+\frac{\h^2}{2v^2}\right)
\Bigg[ \ttw A_{\mu\nu}Z^{\mu\nu} \nn\\&&+\frac{c_{2\theta_W}}{2\ctw^2}  Z_{\mu\nu}Z^{\mu\nu}+ W^+_{\mu\nu}W^{-\mu\nu}
\Bigg]\, .
\label{aazz}
\eea
The first term of \eq{aazz} defines another BSM primary effect: its  contribution to the  $hZ\gamma$ coupling, 
that is constrained by $h\to Z\gamma$ searches. 
Similarly, taking the  $SU(3)_c$ coupling $g_s(\hf)=g_s(1+\boldsymbol{\kappa_{GG}}{\hf^2}/{v^2})$, one obtains 
\be\label{hggluon}
\Delta{\cal L}^h_{{GG}}=\boldsymbol{\kappa_{GG}}\left(\frac{\h}{v}+\frac{\h^2}{2v^2}\right)G^A_{\mu\nu}G^{A\, \mu\nu}\, ,
\ee
whose first term  modifies the $hGG$ coupling  measured in $GG\to h$  \cite{Atalscouplings}, that leads also to 
a per-mille bound on $\boldsymbol{\kappa_{GG}}$~\cite{Pomarol:2013zra}.
Also, from  $Y_f(\hf)$ and $\lambda_h(\hf)$,  we   obtain 
\bea
\Delta{\cal L}^h_{ff}\!\!&=\!\!&\boldsymbol{\delta g^h_{ff}}\Big(\!\h \bar  f_L f_R+\textrm{h.c.}\!\Big)\!\!
 \left(1+\frac{3\h}{2v}+\frac{\h^2}{2v^2}\right)\, ,
\nn\\
\Delta{\cal L}_{3h}\!\!&=\!\!&\boldsymbol{\delta g_{3h}}\, \h^3 \!\left(\!1+\frac{3\h}{2v}+\frac{3\h^2}{4v^2}+\frac{\h^3}{8v^3}\!\right)\, ,
\label{hpe1}
\eea
 with  $\boldsymbol{\delta g^h_{ff}}=-\frac{v}{\sqrt{2}}\partial_{\h} Y_f(\hf)$
and $\boldsymbol{\delta g_{3h}}=-v^2\partial_{\h} \lambda_h(\hf)$,
whose BSM primary effects are respectively the contributions to the $hff$ and  $h^3$ interactions.
Finally, from $Z_h(\hf)$ we obtain,   by going to the canonical basis, the BSM-effect
\be
\Delta{\cal L}_{VV}^h=\boldsymbol{\delta g^h_{VV}} \Bigg[
\h
\left( \!W^{+\mu} W^-_\mu+ \frac{Z^\mu Z_\mu}{ 2 \ctw^2}\!\right) +\Delta_{h}\Bigg]\, ,
\label{hpe2}
\ee
where
\bea
\Delta_h&=&\left( \!W^{+\mu} W^-_\mu+ \frac{Z^\mu Z_\mu}{ 2 \ctw^2}\!\right)\Big(\frac{2 \h^2}{v}+\frac{4 \h^3}{3 v^2}+ \frac{\h^4}{3 v^3}   \Big)\nn\\
&+&\frac{m_h^2 }{12 m^2_W} \left(\frac{\h^4}{v}+\frac{3\h^5}{4 v^2}+\frac{\h^6}{8v^3}\right)
\nn\\  
&+&\frac{m_f }{4m^2_W}\left(\frac{\h^2}{v}+\frac{\h^3}{ 3 v^2}\right)\big(\bar f_L f_R+\textrm{h.c.}\big)\, ,
\eea
$\boldsymbol{\delta g^h_{VV}}=m_W^2 \partial_{\h} Z_h(\hf)$ and we have redefined $\boldsymbol{\delta g_{3h}} \to \boldsymbol{\delta g_{3h}} -m_h^2/(4 m_W^2)\boldsymbol{\delta g^h_{VV}}$ in \eq{hpe1} to eliminate a contribution to the $h^3$ coupling  from $Z_h(\hf)$.
 The first term of \eq{hpe2}  gives a contribution to the  custodial-preserving coupling $hVV$ $(V=Z,W)$ and determines another BSM primary effect. This  coupling can be measured, for example, in $WW\to h$.

It is important at this point to stress that, by construction,  the different $\Delta {\cal L}_{i}$ 
 are \emph{orthogonally}  projected into  different  BSM primary effects, and none of the terms in  a given $\Delta {\cal L}_{i}$ 
contributes to other   BSM primaries that is not its own ({\it e.g.},  no term in $\Delta {\cal L}^h_{{\gamma\gamma}}$ contributes to the $hZ\gamma$,  $hGG$, $hff$, $h^3$ or $hVV$ coupling).
The additional terms in each $\Delta {\cal L}_{i}$, beyond the BSM primary effect, 
tell us what  physical processes are not independent and are instead correlated with the BSM primaries.
This  can be useful if  a departure from the SM predictions is observed:
for example, if only a deviation in    $h\to \gamma\gamma$ is  measured, 
 \eq{gamgam}  tells us that there must also be departures in $h\to ZZ/WW$. Alternatively, if no deviations from the SM are found in the BSM primary effects, these relations can be used to put constraints on the size of the other terms in $\Delta {\cal L}_{i}$.

Having presented all possible interactions   achieved by 
an $\hf$-dependent shift in the  SM parameters, we
study next the set of possible (CP-conserving) BSM contributions  that can lead to departures from gauge-coupling universality.
How many  effects of this type can we have?
Since  EM and $SU(3)_c$  must be  unbroken,   only the $W$ and $Z$ couplings can receive deviations from the SM. 
Assuming for simplicity family universality  we have, in principle,
  9 gauge-boson couplings to fermions  (the $Z$  couplings to $e_{L,R}$, $\nu_L$, $u_{L,R}$, $d_{L,R}$ and the  $W$ couplings to $e_L\nu_L$ and $u_Ld_L$),   the $Z$ coupling to $\hf$, \footnote{\label{foot}We do not independently count the $W$ coupling to $\hf$,  since  this is equivalent to considering  the custodial-preserving combination 
  $\hf^2 (W^\mu W_\mu+Z^\mu Z_\mu/2\ctw^2)$ that  has already been accounted for in \eq{hpe2}.}
and   triple-gauge (TGC) and quartic-gauge (QGC) self-couplings.
We must however keep in mind that  not all  deviations in these couplings are independent from each other, since 
 a  linear combination of all these  corresponds to the universal shift of~\eq{delatastw}. 

Let us first  look at the 10  gauge-boson couplings to fermions and to the Higgs-field $\hf$.
In the  gauge eigenstate basis, corrections to these couplings  arise from 
  the  $\hf$-dependent interactions
\begin{gather}
\hf^2 V_\mu^a J^{\mu\, a}_f\ ,\ \  
\hf^2\eta^a V_\mu^aJ^\mu_{L\, f}\ ,\ \  
\hf^2\eta^a V_\mu^a  J^\mu_{R\, f}\ , \label{vff1} \\
\hf^4\eta^a\eta^b V^a_\mu V^{\mu\, b}\ ,\ \ \ 
\hf^4 \eta^a\eta^b V_\mu^a     J^{\mu\, b}_{L\, f}\, ,\ \ \ 
\label{vff2}
\end{gather}
 where  $V^a_\mu\equiv W^a_\mu- t_{\theta_W}\delta^{a3} B_\mu$ (to preserve EM) and,
 to make the  global  $SU(2)_L$  properties manifest, 
we have separated the  interactions in which the Higgs enters as a  singlet, $\hf^2$, or as a triplet,
\be
\hf^2\eta^a\in H^\dagger \sigma^a H\ ,\ \ \ {\rm with}\  \ \eta^a\equiv(0,0,1)\, .
\ee
The lepton currents are $J^{\mu\, a}_f=\bar L_L\sigma^a\gamma^\mu L_L$,
$J^\mu_{L\, f}=\bar L_L\gamma^\mu L_L$ and 
$J^\mu_{R\, f}=\bar e_R\gamma^\mu e_R$, and similarly for quarks. All  terms of \eq{vff1}  
can arise from ${\cal L}_6$ built as  products of fermion currents  and Higgs currents, these latter being,
in the unitary gauge, 
\be
iH^\dagger \lra D_\mu H=   -  g\frac{ \hf^2}{2}  \eta^a V^a_\mu \ ,\ 
i H^\dagger \sigma^a \lra D_\mu H=  g \frac{ \hf^2}{2}  V^a_\mu \, .
\label{vfrom}
\ee
Similarly, the first term of  \eq{vff2} can  arise from a dimension-6 operator built by  squaring  the first Higgs current of \eq{vfrom}.
On the other hand, the second term in \eq{vff2}, containing four Higgs, can only arise from a dimension-8 operator and  can  then be neglected.
This has the implication that, at the leading order ($\mathcal{L}_6$),
BSM-effects in  the  $W$ couplings are not independent from those in  the   $Z$ couplings.

There are many ways to connect the   BSM-effects of \eq{vff1}   to experiments.  
Since the best constraints on $Vff$ vertices come from measurement of the couplings at the $Z$-pole by LEP,
it is convenient to parametrize the    effects of \eq{vff1} as modifications of the $Z$ couplings to fermions:
\begin{align}\label{shiftl}
&\Delta{\cal L}_{ee}^V=\boldsymbol{\delta g^Z_{eR}} \frac{\hf^2}{v^2} Z^\mu \bar e_R\gamma_\mu e_R\\
&+\boldsymbol{\delta g^Z_{eL}}\frac{\hf^2}{v^2} \left[Z^\mu \bar e_L\gamma_\mu e_L-\frac{c_{\theta_W}}{\sqrt{2}}(W^{+\mu}\bar{\nu_L}\gamma_\mu e_L +\textrm{h.c.}) \right]\nonumber\\
&+\boldsymbol{\delta g^Z_{\nu L}}\frac{\hf^2}{v^2} \left[Z^\mu \bar \nu_L\gamma_\mu \nu_L+\frac{c_{\theta_W}}{\sqrt{2}}(W^{+\mu}\bar{\nu_L}\gamma_\mu e_L +\textrm{h.c.}) \right]\, ,\nn
\end{align}
for leptons, and similarly for quarks:
\begin{align}
&\Delta{\cal L}_{qq}^V=\boldsymbol{\delta g^Z_{uR}}\frac{\hf^2}{v^2} Z^\mu \bar u_R\gamma_\mu u_R+
\boldsymbol{\delta g^Z_{dR}}\frac{\hf^2}{v^2} Z^\mu \bar d_R\gamma_\mu d_R\nn \\
&+\boldsymbol{\delta g^Z_{dL}}\frac{\hf^2}{v^2} \left[Z^\mu \bar d_L\gamma_\mu d_L-\frac{c_{\theta_W}}{\sqrt{2}}(W^{+\mu}\bar{u}_L\gamma_\mu d_L +\textrm{h.c.}) \right]\nonumber\\
&+\boldsymbol{\delta g^Z_{uL}}\frac{\hf^2}{v^2} \left[Z^\mu \bar u_L\gamma_\mu u_L+\frac{c_{\theta_W}}{\sqrt{2}}(W^{+\mu}\bar{u_L}\gamma_\mu d_L +\textrm{h.c.}) \right]\, .
\label{shiftq}
\end{align}
Notice that, as discussed above,  modifications to the $W$ couplings are explicitly related to modifications to the $Z$ couplings.

It remains  to consider the independent effect of the  first term of \eq{vff2}.  
We   consider  it in      the following linear combination 
(that includes also terms of \eq{vff1}):
\begin{align}
\label{fermioncouplingscombo} 
&-\boldsymbol{\delta g_1^Z} c^2_{\theta_W}\frac{\hf^2}{v^2}\Bigg[ 
 \frac{g^2\hf^2}{2}\Big(W_\mu^+ W^{-\mu}+
 \frac{c_{2\theta_W} }{2\ctw^4}Z_\mu Z^\mu\Big)\nn\\
 &+ g (W^-_\mu J^\mu_- +\textrm{h.c.})
+\frac{gc_{2\theta_W}}{\ctw^3} Z_\mu J^\mu_Z +2 e\ttw  Z_\mu J_{em}^\mu
\Bigg]\, ,
\end{align}
 where  $J_Z^\mu=  J_3^\mu{-}s_{\theta_W}^2 J_{em}^\mu$.
Why this particular combination? 
This is    obtained  by performing the EM-preserving shift
${s^2_{\theta_W}}\to s^2_{\theta_W}(1+ 2\boldsymbol{\delta g_1^Z}c ^2_{\theta_W}\hf^2/v^2)$ (keeping $e$ constant),
 {\it only } in the   SM  gauge-couplings of the  fermions  and  $\hf$. 
In the vacuum $\hf=v$, 
the effects in \eq{fermioncouplingscombo} 
can only
be probed as a relative difference of $\stw$ as measured in the  fermion and  $\hf$ sector ($Vff$ couplings and gauge-boson masses), with respect to the value
 as measured in interactions involving gauge bosons only. Therefore it requires the knowledge of  TGC/QGC.
  Indeed, by  field redefinitions, the non-Higgs physics part of \eq{fermioncouplingscombo} can be rewritten    
as a contribution to the   $VWW$ coupling ($g^V$) and  $VV'WW$ coupling ($g^{VV'}$) only. This  explicitly gives
\be\label{TGCQGC}
\boldsymbol{\delta g_1^Z}=\frac{\delta g^{Z}}{ g^{Z}_{SM}}=
\frac{\delta g^{WW}}{2 \ctw^2 g^{WW}_{SM}}=\frac{\delta g^{ZZ}}{2 g^{ZZ}_{SM}}
=\frac{\delta g^{\gamma Z}}{ g^{\gamma Z}_{SM}}\, .
\ee
where $\boldsymbol{\delta g_1^Z}$ has been chosen to
match  the TGC definition of Ref.~\cite{Hagiwara:1986vm}.
\eq{fermioncouplingscombo} gives  however  also a contribution to the custodial-preserving $hVV$ coupling 
that defines one of our BSM primaries,  $\boldsymbol{\delta g^h_{VV}}$.
To eliminate this, we  redefine $\boldsymbol{\delta g^h_{VV}}\to \boldsymbol{\delta g^h_{VV}}+g^2v\boldsymbol{\delta g_1^Z}\ctw^2$ in \eq{hpe2},  that gives an  extra contribution 
 proportional $\boldsymbol{\delta g_1^Z}$  to be added to    \eq{fermioncouplingscombo}.
 The final result is    
\begin{align}
\label{fermioncouplings} 
&\!\Delta{\cal L}_{\!g_1^Z}\!= \!\boldsymbol{\delta g_1^Z}\! \Bigg[  ig c_{\theta_W}\!\Big(\!Z^\mu (W^{+ \nu} W^{-}_{\mu\nu}\!-\!\textrm{h.c.}\!)\!+\!Z^{\mu\nu}W^+_\mu W^-_\nu\!\Big)
 \nn\\
 -&2gc_{\theta_W}^2\!\frac{\h}{v}
\Bigg(
 W^-_\mu J^\mu_- \!+\!\textrm{h.c.}
\!+\!\frac{c_{2\theta_W}}{\ctw^3} Z_\mu J^\mu_Z 
\!+\frac{2\stw^2}{\ctw}  Z_\mu J_{em}^\mu
\Bigg)
 \nn\\
&\,\times\!\!\left(1+\frac{\h}{2v}\right)+\frac{e^2v}{2\ctw^2}\h Z_\mu Z^\mu
+g^2c_{\theta_W}^2v\, \Delta_h
 \\
 \!-\!
 &g^2c_{\theta_W}^2\!\Big(\!W_\mu^+ W^{-\mu}+
 \frac{c_{2\theta_W} }{2\ctw^4}Z_\mu Z^\mu\!\Big) \!\!\left(\frac{5\h^2}{2}+\frac{2\h^3}{v}+\frac{\h^4}{2v^2}\!\right)\!\!
\Bigg]\nn\, .
\end{align}

The interesting property of our parametrization  Eqs.~(\ref{shiftl},\ref{shiftq},\ref{fermioncouplings}) is the following.
Since BSM-effects  to   SM propagators can always be eliminated through the  equations of motion (EOM), 
there is a one to one correspondence  between each of the  $\boldsymbol{\delta g^Z_{f}}$ of Eqs.~(\ref{shiftl},\ref{shiftq})
  and the corresponding $\Gamma(Z\to ff)$ partial-width measured at LEP1~\cite{ALEPH:2005ab}.\footnote{This is  true in the limit in which $m_f$ is neglected, so that interference with  dipole-type BSM contributions vanishes.}
Therefore all the 7 parameters $\boldsymbol{\delta g^Z_f}$ 
can be   bounded at the per-mille level by $Z$ decay-widths and asymmetries 
at LEP1,\footnote{These measurements can also be combined with the measurement of the $Wud$ coupling as extracted  
from $K$ and $\beta$-decays in  combination  with information on 4-fermion interactions from the LHC \cite{Pomarol:2013zra}.}
 using  $\alpha_{em}$, $m_W$ and $m_Z$ as the  SM input parameters. 
This latter choice 
makes the phenomenological analysis particularly  transparent, since these input parameters receive no corrections 
from  BSM-effects.\footnote{\label{footnotexx}Four-fermion interactions in $\mathcal{L}_6$ (which have no direct relation with Higgs physics and can therefore be studied  separately) affect the value of $G_F$ as extracted through the measurement of $\mu$-decay. For this reason the  traditional choice of using $G_F$ to fix one input parameter  is less convenient than the one we propose here.}
 On the other hand, $\boldsymbol{\delta g_1^Z}$ is constrained by TGC measurements at LEP2 \cite{Schael:2013ita} and LHC.
Although a  global analysis on the ${\cal L}_6$  contributions to TGC, using all existing data, does not exist yet, 
we expect that bounds on $\boldsymbol{\delta g_1^Z}$ can reach at present the per-cent level~\cite{TGCHouches,TGCtoappear}.
\eq{fermioncouplings}  gives also contributions to Higgs physics, such as a custodial-breaking  $hVV$ coupling
or new effects to $h\to V\bar f f$ that could  be  also  used to constrain $\boldsymbol{\delta g_1^Z}$.
We believe however that when rigorous analyses of both TGC
and Higgs physics  (on the lines of Ref.~\cite{IsidoriTrot}) will be available, TGC constraints will always outdo Higgs physics ones~\cite{TGCtoappear}, so that our parametrization will remain the most  convenient.

At this point the reader might be surprised by the unorthodox choice of abandoning the traditional parametrization 
of BSM-effects that includes the $S$ and $T$-parameters \cite{Peskin:1990zt}, in favor of a description in terms of vertex corrections as given in Eqs.~(\ref{shiftl},\ref{shiftq}).   
As already explained, we consider  the latter  more suited to compare with the experimental data.
Nevertheless, if desired,
two of the parameters  of Eqs.~(\ref{shiftl},\ref{shiftq}), {\it e.g.}, $\boldsymbol{\delta g^Z_{eL}}$ and $\boldsymbol{\delta g^Z_{\nu L}}$ \cite{Pomarol:2013zra}, can always  be traded  with
\begin{align}\label{SandT}
&\Delta{\cal L}_{\hat S}=\boldsymbol{\hat S}\, \frac{g\stw^2}{\ctw^3} \frac{\hf^2}{v^2} \,Z_\mu\left[ J^\mu_Z-c_{\theta_W}^2J_{em}^\mu+\frac{ g}{c_{\theta_W}} \frac{\hf^2}{4}Z^\mu\right]\, ,\nn\\
&{\Delta \cal L}_T=-\frac{\boldsymbol{\hat T}}{2}\frac{ \hf^4}{v^4} m_Z^2 Z^\mu Z_\mu\, .
\end{align}
It is clear that Eqs.~(\ref{SandT}) can arise from  linear combinations of  terms in \eq{vff1} and  \eq{vff2}, and therefore
gives  a different parametrization from  Eqs.~(\ref{shiftl},\ref{shiftq}).
To see that indeed $\boldsymbol{\hat S}$ is the $S$-parameter (we use the $S,T$ normalization of Ref.~\cite{Barbieri:2004qk}), one must realize  that  the terms in  brackets in
$\Delta{\cal L}_{\hat S}$ are proportional to $J_Y^\mu$,
and therefore (by the EOM)   can be written as $\Delta{\cal L}_{\hat S}=-\boldsymbol{\hat S}(s_{\theta_W}/c^2_{\theta_W})(\hf^2/v^2) \,Z_\mu\partial_\nu B^{\mu\nu}$. After partial integration, this gives
\begin{align}
\Delta{\cal L}_{\hat S}= &{ -\frac{\boldsymbol{\hat S}}{2}}\Big[{t_{\theta_W}}\, \frac{\hf^2}{v^2} B_{\mu\nu}\hat W_3^{\mu\nu} -{t_{\theta_W}^2}\, \frac{\hf^2}{v^2} B_{\mu\nu}B^{\mu\nu}\nn\\
&{ -2}\, \frac{\partial_\nu \hf^2}{v^2} B_{\mu\nu}(t_{\theta_W}W_3^{\mu}-t_{\theta_W}^2B^{\mu})\Big]\, .
\label{Sexpanded}
\end{align}
For $\hf=v$,  
the first term
coincides with the  $\hat S$ parameter~\cite{Barbieri:2004qk},
while   the second term merely represents a redefinition of the SM parameter $g^\prime$, and the last one $\propto\partial_\nu \hf^2$ vanishes.

Apart from \eq{vff1}, there can also be   BSM-induced interactions between fermion and   gauge-bosons 
of a different  type than those in the SM. 
  These  include  couplings of  $W$ to right-handed quarks 
and dipole-type interactions, that we parametrize as 
\begin{align}
&\Delta {\cal L}^W_{R}=\boldsymbol{\delta g^{W}_R} \frac{\hf^2}{v^2}W^+_{\mu}\bar u_R\gamma^\mu  d_R
+\textrm{h.c.}\, ,\label{dipoles}\\
&\Delta {\cal L}^V_{\rm dipole}=\frac{eY_q\hf}{m_W^2} \Big[
\boldsymbol{\delta \kappa^G_{q}}\, \frac{g_s}{e}  \bar q_LT^A \sigma^{\mu\nu}q_R G^A_{\mu\nu}\nn\\
&+\boldsymbol{\delta \kappa^A_{q}}   (  T_3 \bar{q}_L \sigma^{\mu\nu} q_R A_{\mu\nu}+\frac{\stw}{\sqrt{2}}   \bar{u}_L \sigma^{\mu\nu} d_R W_{\mu\nu}^+)\nn\\
&+\boldsymbol{\delta \kappa^Z_{q}}   (T_3  \bar{q}_L \sigma^{\mu\nu} q_R  Z_{\mu\nu}+\frac{\ctw}{\sqrt{2}}   \bar{u}_L \sigma^{\mu\nu} d_R W_{\mu\nu}^+)+\textrm{h.c.}\Big]\, ,\nn
\end{align}
 for quarks $q=u,d$, where the coefficients are assumed to be real and $T_3$ denotes weak isospin
 (and similarly for leptons).
 Note that the dipole interactions  with $W$ are  not independent from those of  $A$ and $Z$, as the term
that splits these dipole interactions, $ \hf^2 \eta_a W^{a}_{ \mu\nu}\hf \bar{q}_L \sigma^{\mu\nu} q_R   $, arises  at 
 dimension-8.

Let us now move   to TGC and QGC. At $O(p^4)$
there are 4 possible CP-conserving  TGC couplings and 5  QGC \cite{Reuter:2013gla}, but  not all  can arise from ${\cal L}_6$.
We already encountered one with the same Lorentz structure as in the SM: $\Delta{\cal L}_{  g_1^Z}$ that led to
\eq{TGCQGC}.
Other contributions
could in principle arise from ${\cal L}_6$ operators containing covariant derivatives and/or field-strengths.
However, by integration by parts and using the EOM, one can reduce them to dimension-6 operators with only field-strengths  \cite{Grzadkowski:2010es}.\footnote{Dimension-6 operators 
with Higgs and derivatives can always be written, using integration by parts, as  operators of the type 
$H^\dagger D_\mu  H D_\nu F^{\mu\nu}$
and $H^\dagger D_\mu D_\nu HF^{\mu\nu}$. The former reduces to  \eq{vff1} and  \eq{vff2}   by the gauge-boson EOM, while  the latter can, by means of $[D_\mu, D_\nu]=F_{\mu\nu}$, be written as an operator made of field-strengths only. Operators built with  \eq{vfrom}  contributing to TGC/QGC are always of dimension larger than six.}
 The only operators  at $O(p^4)$, made of  field-strengths   and contributing to EWSB, are
\be\label{SU}
\hf^2\eta^a W^a_{\mu\nu}B^{\mu\nu}\, ,\ \ \ 
\hf^4\eta^a\eta^b W^a_{\mu\nu}W^{b\, \mu\nu}\, .
\ee
The second  one involves four Higgs and  cannot arise from dimension-6 operators, 
while the first one  gives
\begin{align}
&\hf^2\eta^a W^a_{\mu\nu}B^{\mu\nu}=\label{fs}\\
&\quad \hf^2\left[\hat  W^3_{\mu\nu} B^{\mu\nu} +2 ig\ctw W^-_{\mu} W^{+}_{\nu}(A^{\mu\nu}-\ttw Z^{\mu\nu})\right]\,  .\nn
\end{align}
The  second term clearly contains  a new dipole-type TGC for the $W$,
that can be identified  with  $\delta \kappa_ \gamma$ of  Ref.~\cite{Hagiwara:1986vm}.
Since  \eq{fs} also contains contributions to   other BSM primaries (such as $\hat S$ and $h\to \gamma\gamma,Z\gamma$), we must arrange a linear combination
that does not project into them.
From the relation
\begin{align}
\ttw\frac{\hf^2}{2 v^2}W^3_{\mu\nu}B^{\mu\nu}=&
-\frac{\Delta{\cal L}_{\hat{S}}}{\boldsymbol{\hat{S}}}+\stw^2\frac{\Delta{\cal L}^h_{{\gamma \gamma}}}{\boldsymbol{\kappa_{\gamma\gamma}}}+c_{2 \theta_W}\frac{\Delta{\cal L}^h_{{Z \gamma}}}{\boldsymbol{\kappa_{Z\gamma}}} \nn\\
&
-\frac{\Delta{\cal L}_{\kappa_\gamma}}{\boldsymbol{\delta \kappa_\gamma}}\, ,
\end{align}
we have that
\begin{align}
 \Delta{\cal L}_{\kappa_\gamma}&=\frac{\boldsymbol{\delta \kappa_{\gamma}}}{v^2}\Big[ie \hf^2(A_{\mu\nu}- t_{\theta_W}Z_{\mu\nu})W^{+\mu} W^{-\nu}\label{kapgam}\\
&+Z_\nu \partial_\mu \hf^2 (t_{\theta_W}A^{\mu \nu}-t_{\theta_W}^2 Z^{\mu\nu})+\frac{(\hf^2-v^2)}{2}\nn\\
\times&\Big(\ttw  Z_{\mu\nu}A^{\mu\nu}+\frac{c_{2 \theta_W}}{2 \ctw^2} Z_{\mu\nu}Z^{\mu\nu}+W^+_{\mu\nu}W^{-\mu\nu}\Big)\Big]\, ,\nn
\end{align}
gives  us the combination that we were looking for: it projects into  
a  new BSM primary effect,
the TGC $\delta \kappa_ \gamma$  \cite{Hagiwara:1986vm},
 but not into previous ones.

The list  of  the $\hf$-dependent interactions presented so far is in principle not  complete.
We could also have dimension-6 interaction terms of the following type:
$(\partial_\mu \hf)^2 V^\nu V_\nu$,  $\partial_\mu\hf V^\mu V_ \nu V^\nu$,
$(\partial_\mu \hf) V_\nu \bar{f}_L \sigma^{\mu \nu} f_R$ and  $\hf V_\mu V^\mu \bar{f}_L{f}_R$ (other ones can be reduced to previous ones by redefinitions  or partial integration). 
Nevertheless, it is easy to check that none of these terms can arise from operators  in ${\cal L}_6$  independent  from the ones introduced before.

The  list of CP-even independent effects  is completed by 4-fermion interactions $\Delta {\cal L}_{\rm 4f}$ and $O(p^6)$ interactions, both of which cannot contain the Higgs-field $\hf$ if they arise from dimension-6 operators (they preserve the SM symmetries). 
The relation  of 4-fermion interactions with experiments is straightforward and they do not interfere with the observables determining  our previous BSM primaries  (see footnote~\ref{footnotexx}). The complete list can be found in
Ref.~\cite{Grzadkowski:2010es}.
On the other hand, we have two $O(p^6)$ interactions,
   $\epsilon_{abc}W^{a\, \nu}_{\mu}W^{b}_{\nu\rho}W^{c\, \rho\mu}$ and $\epsilon_{ABC}G^{A\, \nu}_{\mu}G^{B}_{\nu\rho}G^{C\, \rho\mu}$,     since   terms with only two  gauge field-strengths can be eliminated by EOM.
In the physical basis these are given by
\bea
\Delta {\cal L}_{\lambda_\gamma}&=&\frac{i\boldsymbol{\lambda_\gamma}}{m^2_W}\left[ (eA^{\mu\nu}+g\ctw Z^{\mu\nu})
  W^{- \rho}_{\nu}  W^{+}_{\rho \mu}\right]\, ,\nn\\
\Delta {\cal L}_{3G}&=&\frac{\boldsymbol{ \kappa_{3G}}}{m^2_W}g_s\epsilon_{ABC}G^{A\, \nu}_{\mu}G^{B}_{\nu\rho}G^{C\, \rho\mu}\, .
 \label{lamgam} 
\eea
The former projects onto the TGC observable $\lambda_\gamma$
defined in  Ref.~\cite{Hagiwara:1986vm}
 and   can interfere with   the extraction of  $\boldsymbol{\delta g_1^Z}$ and $\boldsymbol{\delta\kappa_\gamma}$ from $ff\to WW$ \cite{TGCHouches}.

For completeness, we also briefly comment on CP-violating interactions $\Delta {\cal L}_{\rm CPV}$, beyond those  in $\Delta {\cal L}_{\rm 4f}$: these can easily be obtained from the CP-even ones of
\eq{hpe1} and  \eq{dipoles}  by
$\boldsymbol{\delta g^h_{ff}}\to i\boldsymbol{\delta\tilde g^h_{ff}}$,
$\boldsymbol{\delta g^W_{R}}\to i\boldsymbol{\delta \tilde g^W_{R}}$,
$\boldsymbol{\delta \kappa^V_{q}}\to i\boldsymbol{\delta\tilde \kappa^V_{q}}$, 
and by substituting one of the field strengths 
with its dual ($V_{\mu \nu} \to \tilde{V}_{\mu \nu}$) in each term of  Eqs.~(\ref{gamgam}), (\ref{aazz}), (\ref{hggluon}), (\ref{kapgam})  and (\ref{lamgam}).

 Having written  the  BSM effects as  interaction-terms  allows us an easy estimate of their coefficients.
 The BSM primaries $\boldsymbol{\delta g_i}$
 (as well as 4-fermion interactions) are expected to scale as  \cite{Elias-Miro:2013mua}
 \be
\frac{\boldsymbol{\delta g^{h}_{ff}}}{Y_f}\sim
\frac{\boldsymbol{\delta g_{3h}}}{\lambda_h v}\sim
\frac{\boldsymbol{\delta g^{h}_{VV}}}{gm_W}\sim
\frac{\boldsymbol{\delta g^{\!Z,W}_{\!f\!,R}}}{g}\sim
\boldsymbol{\delta g^Z_{1}}\sim
\frac{g_*^2v^2}{\Lambda^2}\, ,
 \ee
 where $g_*$ denotes a generic  BSM coupling. Our operator expansion makes sense if ${g_*^2v^2}/{\Lambda^2}\ll 1$. 
On the other hand, the BSM primaries $\boldsymbol{\delta \kappa_i}$ and  $\boldsymbol{\lambda_\gamma}$, associated to interactions carrying extra derivatives or gauge-fields, are expected to scale as $ g^2v^2/\Lambda^2$, where $g$ is the corresponding    SM gauge-coupling. 
 This means that for strongly coupled theories in which $g_*\sim 4\pi$, the BSM primaries $\boldsymbol{\delta g_i}$
are potentially  more important.
Also
in renormalizable weakly-coupled theories  $\boldsymbol{\delta \kappa_i}$ and  $\boldsymbol{\lambda_\gamma}$ are  suppressed by at least a  one-loop factor \cite{Elias-Miro:2013mua}.
 In the particular case of   TGC/QGC,  we have that $\boldsymbol{\delta g^Z_{1}}$ could  potentially  give the most sizeable effects, having an important impact on high-energy processes  if $g_*$ is large \cite{inprep}.

In summary, the Lagrangian up to dimension-6 operators can be written as ${\cal L}={\cal L}_{\rm SM}+\Delta{\cal L}_{\rm BSM}$ with
\begin{align}
&\!\!\!\!\!
\Delta{\cal L}_{\rm BSM}=
\Delta{\cal L}^h_{{\gamma \gamma}}+\Delta{\cal L}^h_{{Z \gamma}}+\Delta{\cal L}^h_{{GG}}
+\Delta{\cal L}^h_{ff}
+\Delta{\cal L}_{3h}\nn\\
&\!\!\!\!\!\!
+\Delta{\cal L}^h_{VV}
+\Delta{\cal L}^V_{ee}+\Delta{\cal L}^V_{qq}+\Delta {\cal L}_{R}^W+\Delta {\cal L}^V_{\rm dipole}\nn\\
&\!\!\!\!\!\!
+\Delta{\cal L}_{g_1^Z}+\Delta{\cal L}_{\kappa_\gamma}
+\Delta {\cal L}_{\lambda_\gamma}+\Delta {\cal L}_{3G}
+\Delta {\cal L}_{\rm 4f}+\Delta {\cal L}_{\rm CPV}\, .\label{Ltot}
\end{align}
The first term of each  $\Delta{\cal L}_i$  gives a  BSM primary effect  which sets, at present or in the near future, the most compelling constraint on the coefficient of $\Delta{\cal L}_i$.
Notice that  $\Delta{\cal L}_{\rm BSM}$ only includes interaction terms, as  BSM
contributions to  propagators  have been eliminated through the EOM, making the connection between BSM primaries and observables particularly transparent.
The main predictions from 
 \eq{Ltot} are:
the $Wff$ and $Zff$ vertices are related (Eqs.~(\ref{shiftl},\ref{shiftq})); 
the $W$ dipole-type interaction for the fermions are related to those of $A$ and $Z$ (\eq{dipoles});
 there are only 3 types of CP-conserving TGC,  characterized by $\delta g_1^Z$, $\delta\kappa_\gamma$ and 
 $\lambda_\gamma$   \cite{Hagiwara:1986vm}, while  QGC are related to them  by  Eqs.~(\ref{TGCQGC},\ref{lamgam});
there are only 8  Higgs BSM primary effects (for one family) \cite{Elias-Miro:2013mua} while all other Higgs interactions  ({\it e.g.} effects to $h\to Vff$ or custodial-breaking $hVV$ couplings)
 are related to  BSM primaries~\cite{Pomarol:2013zra,inprep}.

 We are  thankful to Eduard Masso, for discussions and precious comments. FR  acknowledges support from the Swiss National Science Foundation, under the Ambizione grant PZ00P2 136932.
The work of AP has  been  supported by  the grants FPA2011-25948,  2009SGR894
and the Catalan ICREA Academia Program.


\end{document}